\documentclass{article}

\usepackage{amsmath}
\usepackage{amssymb}
\usepackage{amsfonts}
\usepackage{epic}
\usepackage{epsfig}
\usepackage{subfigure}

\def\N{{\mathbb N}}
\def\Z{{\mathbb Z}}
\def\s{{\bf s}}
\def\q{{q}}

\newcounter{NN}
\newtheorem{conjecture}[NN]{Conjecture}

\begin{document}
\bibliographystyle{plain}
\title{Growth of degrees of integrable mappings}
\author{Peter H.~van der Kamp}
\date{Department of Mathematics and Statistics,\\
La Trobe University, \\
3086, Australia\\[5mm]
date: \today }

\pagestyle{plain} \maketitle

\begin{abstract}
We study mappings obtained as $\s$-periodic reductions of
the lattice Korteweg-De Vries equation. For small $\s\in\N^2$
we establish upper bounds on the growth of the degree of the
numerator of their iterates. These upper bounds appear to be exact.
Moreover, we conjecture that for any $s_1,s_2$ that are co-prime the
growth is $\sim (2s_1s_2)^{-1}n^2$, except when $s_1+s_2 = 4$ where the
growth is linear $\sim n$. Also, we conjecture the degree of the
$n$-th iterate in projective space to be
$\sim (s_1+s_2)(2s_1s_2)^{-1}n^2$.

\end{abstract}

\section{Introduction}
Integrable mappings are characterised by low complexity \cite{A,Ves}.
This idea culminated in the notion of algebraic entropy, introduced
by Viallet and collaborators \cite{BV,FV,HV98}. Low complexity means
vanishing algebraic entropy which corresponds to polynomial growth of
degrees of iterates of the mapping. A first proof of such a polynomial
bound on the degrees was given in \cite{BMR}. In \cite{B} it was proven
that foliation by invariant curves implies zero algebraic entropy.
Examples show that degree growth is a better indication of integrability
than singularity confinement \cite{HV98,HV00}, cf. the discussion in
\cite{OTGR}. Recently the notion has been extended to lattice equations
\cite{TGR,Via} and used to find new integrable models \cite{HV07}.

In practise, one calculates the growth of degrees $d_n$ of the first $n$ iterates
of a mapping. Then one guesses the pattern by fitting the generating
function $g(x)=\sum d_n x^n$ with a rational function $p(x)/q(x)=g(x)$ and the algebraic
entropy $\text{lim}_{n\rightarrow \infty} \text{log}(d_n)/n$ is obtained as the logarithm
of the inverse of the smallest zero of $q(x)$, see \cite{Via}. We present an elementary
method that enables one to derive upper bounds for the growth of degrees. Our formulas
exactly produce all degrees that we have been able to calculate.

\section{Outline}
We will perform $\s$-periodic reductions of the lattice Korteweg-De Vries equation
\begin{equation} \label{kdv}
(u_{l,m}-u_{l+1,m+1})(u_{l+1,m}-u_{l,m+1})=\alpha.
\end{equation}
This corresponds to studying solutions that satisfy the periodicity condition
$u_{l,m}=u_{l+s_1,m+s_2}$. We choose $s_1$ and $s_2\leq s_1$ to be co-prime
natural numbers. Under this assumption the lattice equation reduces to a
single ordinary difference equation (O$\Delta$E) of order $\q := s_1+s_2$ (or, a
$\q$-dimensional mapping).
For background on periodic reductions we refer to \cite{K,QCPN}. There are
$\q$ initial values, which we denote by $x_1,x_2,\ldots,x_\q$.
The O$\Delta$E, or the mapping, can be used to generate a solution $x_{n\in\Z}$,
which are rational functions in the initial values.

One aim is to find a formula for the degree
of the numerator (or denominator) of $x_n$, as a function of $n$. We set $x_n=a_n/b_n$,
and derive a system of two O$\Delta$Es for $a_n$ and $b_n$, which are polynomials
in the initial values. By choosing $b_n=1$ for $n=1,2,\ldots,\q$ the degree (i.e., total degree in the variables $x_1=a_1,\ldots,x_\q=a_\q$) of the numerator of $x_n$ is given by
$d^a_n-d^g_n$. Here $d^p_n$ denotes the $d$egree of a polynomial $p_n$, and $g_n$ is
the greatest common divisor $g_n=\text{gcd}(a_n,b_n)$.
First we obtain a recursive formula for $d^a_n=d^b_n+1$. Then we look at the growth of $g_n$.
After a number of iterates a miracle occurs: any divisor of $b_n$ will divide $g_{n+\q}$ ($\q\neq 4$).
This statement has been verified for a range of periodicities $\s$, but seems to be difficult
to prove in general. Next, we find a recurrence formula for the growth of the multiplicities of
divisors: a divisor of $g_n$ divides $g_{n+i}$ with multiplicity $t_i$, where $t$ is
an integer sequence satisfying a linear recurrence relation. We define a new set of polynomials
$c_n=b_n/f$, where $f$ is the product of all divisors
of $b_{i<n}$ with the right multiplicities as given by the integer sequence $t$.
Multiplying by $f$ (which is a product $c_{i<n}$'s) and taking the degree on both
sides of $c_n f = b_n$, we find that $d^c_n+(d^c\ast t)_n = d^b_n$ where $\ast$ denotes discrete
convolution
\begin{equation} \label{conv}
(d\ast t)_{n+1}=d_1t_n+d_2t_{n-1}+\cdots +d_nt_1.
\end{equation}
Using the recursive formulas for $d^b$ and $t$ we find a recursive formula for
$d^c$, which can be solved to find polynomial growth of degree $2$. Moreover,
we obtain the coefficient of the leading term: $(2s_1s_2)^{-1}$.

We also consider the projective analogues of these mappings. We introduce
homogeneous coordinates and derive a polynomial
mapping in $\q$-dimensional projective space.  Here, the aim is to find a formula
for degree of the components of this mapping. The strategy is very similar as the
above. Once one has a divisor $c_i$ of certain components of the mapping one can derive
a recursive formula for the multiplicities at higher iterates of the mapping. At a
certain point these multiplicities are (miraculously) higher than expected,
after which the growth can be described recursively again. As before, a convolution
formula provides us with a recurrence for the degrees of the divisors. In this
case the degree of the $n$-th iterate is given by the sum $1+d^c_{n-1}+d^c_{n-2}+
\cdots +d^c_{n-\q}$. This growth can also be described recursively and the
leading term is found to be $(s_1+s_2)(2s_1s_2)^{-1}n^2$.

The case $s=(3,1)$ is exceptional. Here the growth is linear $\sim n$, and the mapping
is linearisable. We provide its explicit solution in terms of an interesting sequence
of polynomials, see section \ref{ExCa} and the appendix.

\section{Growth of degrees of rational mappings}
We first illustrate our approach by considering a low dimensional
example, taking $\s=(2,1)$.
\subsection{A low dimensional example}
\begin{figure}[h]
\setlength{\unitlength}{1cm}
\begin{center}
\begin{picture}(5,3)(0,0)
\matrixput(0,0)(1,0){6}(0,1){4}{\circle{.1}}
\put(0,0){\line(0,1){1}} \put(2,1){\line(0,1){1}}
\put(4,2){\line(0,1){1}} \put(0,1){\line(1,0){2}}
\put(2,2){\line(1,0){2}} \put(4,3){\line(1,0){1}}
\put(0,0){\circle*{.1}} \multiput(0,1)(1,0){3}{\circle*{.1}}
\multiput(2,2)(1,0){3}{\circle*{.1}}
\multiput(4,3)(1,0){2}{\circle*{.1}}
\put(0.1,0.1){$x_3$}
\put(1.1,0.1){$x_4$}
\put(2.1,0.1){$x_5$}
\put(3.1,0.1){$x_6$}
\put(4.1,0.1){$x_7$}
\put(5.1,0.1){$x_8$}
\put(0.1,1.1){$x_1$}
\put(1.1,1.1){$x_2$}
\put(2.1,1.1){$x_3$}
\put(3.1,1.1){$x_4$}
\put(4.1,1.1){$x_5$}
\put(5.1,1.1){$x_6$}
\put(2.1,2.1){$x_1$}
\put(3.1,2.1){$x_2$}
\put(4.1,2.1){$x_3$}
\put(5.1,2.1){$x_4$}
\put(4.1,3.1){$x_1$}
\put(5.1,3.1){$x_2$}
\end{picture}
\caption{\label{ivp} Staircase with $(1,2)$-periodic initial values ($x_1,x_2,x_3$), solved to the right.}
\end{center}
\end{figure}
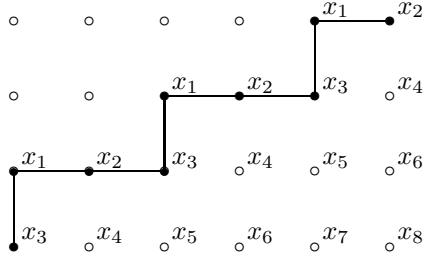

We take initial values $x_1,x_2,x_3$ on a staircase as in
Figure \ref{ivp}.

The $x_n$ are rational functions of $x_1,x_2,x_3,\alpha$
which can be calculated recursively using
\begin{equation} \label{recx}
x_n=P(x_{n-1},x_{n-2},x_{n-3}),
\end{equation}
where $P$ solves equation (\ref{kdv}) for $u_{l+1,m}$,
\begin{equation}\label{P}
u_{l+1,m} = P(u_{l,m},u_{l+1,m+1},u_{l,m+1}) := u_{l,m+1}+\frac{\alpha}{u_{l,m}-u_{l+1,m+1}}.
\end{equation}

We write $x_n=a_n/b_n$. The recurrence (\ref{recx}) yields
the following recurrences for $a,b$:
\begin{align}
a_n&=a_{n-3}w_n-\alpha b_{n-1}b_{n-2}b_{n-3} \tag{5a}\\
b_n&=b_{n-3}w_n \tag{5b}
\end{align}
\addtocounter{equation}{1}
where $w_n=a_{n-2}b_{n-1}-a_{n-1}b_{n-2}$. We choose $b_1=b_2=b_3=1$, so
that $a_n$ and $b_n$ are polynomials in the variables (initial values)
$a_1,a_2$ and $a_3$. Their total degree will be denoted $d^a_n$ and $d^b_n$,
respectively. From (5) it follows that the degrees are at most
\begin{eqnarray*}
d^a_n&=&\text{max}(d^b_{n-1}+d^a_{n-2}+d^a_{n-3},d^a_{n-1}+d^b_{n-2}+d^a_{n-1},
d^b_{n-1}+d^b_{n-2}+d^b_{n-3}),\\
d^b_n&=&\text{max}(d^b_{n-1}+d^a_{n-2}+d^b_{n-3},d^a_{n-1}+d^b_{n-2}+d^b_{n-3}).
\end{eqnarray*}
Given the initial degrees $d^a_n=d^b_n+1=1$ ($n=1,2,3$) we find that
\begin{eqnarray}
d^a_n&=&d^a_{n-1}+d^a_{n-2}+d^a_{n-3}-1, \notag \\
d^b_n&=&d^b_{n-1}+d^b_{n-2}+d^b_{n-3}+1 \label{dbn}
\end{eqnarray}
are upper bounds for the degrees of $a_n$ and $b_n$, and $d^a_n=d^b_n+1$ ($n\in\N$).
The sequence $d^b$ comprises sums of tribonaccci numbers, cf. \cite[seq. A008937]{Slo}.
Certainly, these sequences grow exponentially. However, there will be a lot of cancelations
in $x_n=a_n/b_n$ due to common factors of $a_n,b_n$. We will prove that the degree of the
greatest common divisor
\[
g_n=\text{gcd}(a_n,b_n)
\]
is sufficiently large to ensure that $d^a_n-d^g_n$ grows polynomially.

Suppose that $f^{t_k}$ divides $g_k$ with $k\in\{n-1,n-2,n-3\}$.
Then from (5) it follows that $f^{t_n}$ divides $g_n$,
where
\begin{equation} \label{tn}
t_n=t_{n-1}+t_{n-2}+t_{n-3}.
\end{equation}
We define an integer sequence $t$ by $t_1=t_2=t_3-2=0$ and the above recursion.
Such numbers $t$ are called tribonacci numbers, cf. \cite[seq. A000073]{Slo}.
Thus have the following:
\[
f^2|g_n \Rightarrow f^{t_{3+i}} | g_{n+i}, \ i\in\N.
\]

By direct calculation, using Maple and the recurrences (5), we find
that the polynomial $w_n^2$ divides $g_{n+3}$ ($n>3$). This implies that
$w_n^{t_i}$ divides $g_{n+i}$. We can now write symbolically
\begin{eqnarray}
b_i&=&c_i=1,\ i=1,2,3, \notag \\
b_i&=&c_i,\ i=4,5,6, \notag \\
b_7&=&c_7c_4^2, \notag \\
b_8&=&c_8c_4^2c_5^2=c_8c_4^{t_4}c_5^{t_3}, \notag\\
b_9&=&c_9c_4^4c_5^2c_6^2=c_9c_1^{t_8}c_2^{t_7}c_3^{t_6}c_4^{t_5}
c_5^{t_4}c_6^{t_3}c_7^{t_2}c_8^{t_1}, \notag \\
&\vdots& \notag \\
b_n&=&c_n \prod_{i=1}^{n-1} c_i^{t_{n-i}}, \label{bn}
\end{eqnarray}
which defines polynomials $c_n$. Taking the degree of both sides of
equation (\ref{bn}) we find $d^b_n= d^c_n + (d^c \ast t)_n$ where
$\ast$ denotes discrete convolution, see (\ref{conv}). From this we infer,
using the recurrence for $t$ (\ref{tn}), that
\begin{eqnarray*}
d^b_n - d^c_n &=& d^c_1 t_{n-1} + \cdots + d^c_{n-4} t_4 + d^c_{n-3} t_3 \\
&=& d^c_1 (t_{n-2}+t_{n-3}+t_{n-4}) + \cdots + d^c_{n-4} (t_3+t_2+t_1) + d^c_{n-3} t_3 \\
&=& (d^c\ast t)_{n-1} + (d^c\ast t)_{n-2} + (d^c\ast t)_{n-3} + 2 d^c_{n-3} \\
&=& d^b_{n-1}-d^c_{n-1} + d^b_{n-2}-d^c_{n-2} + d^b_{n-3} + d^c_{n-3}
\end{eqnarray*}
which, using the recursion for $d^b$ (\ref{dbn}), shows that
\[
d^c_n=d^c_{n-1} + d^c_{n-2} - d^c_{n-3} + 1.
\]
Together with $d^c_1=d^c_2=d^c_3=0$ this gives a sequence of quarter-squares,
cf. \cite[seq. A033638]{Slo},
\[
d^c_n= \lfloor \frac{(n-2)^2}{4} \rfloor.
\]
Note that the $c_{i<n}$'s in (\ref{bn}) are divisors of $g_n$. Thus the
quantity $d^a_n-d^g_n$ is bounded from above by $d^b_n+1-(d^c \ast t)_n=d^c_n+1$,
which grows asymptotically $\sim n^2/4$.

\subsection{More general periodic reductions}
\begin{figure}[h]
\setlength{\unitlength}{1cm}
\begin{center}
\begin{picture}(5,3)(0,0)
\matrixput(0,0)(1,0){6}(0,1){4}{\circle{.1}}
\put(0,0){\line(0,1){1}} \put(2,1){\line(0,1){1}}
\put(3,2){\line(0,1){1}} \put(0,1){\line(1,0){2}}
\put(2,2){\line(1,0){1}} \put(3,3){\line(1,0){2}}
\put(0,0){\circle*{.1}} \multiput(0,1)(1,0){3}{\circle*{.1}}
\multiput(2,2)(1,0){3}{\circle*{.1}}
\multiput(4,3)(1,0){2}{\circle*{.1}}
\put(0.1,0.1){$x_4$}
\put(1.1,0.1){$x_6$}
\put(2.1,0.1){$x_8$}
\put(3.1,0.1){$x_{10}$}
\put(4.1,0.1){$x_{12}$}
\put(5.1,0.1){$x_{14}$}
\put(0.1,1.1){$x_1$}
\put(1.1,1.1){$x_3$}
\put(2.1,1.1){$x_5$}
\put(3.1,1.1){$x_7$}
\put(4.1,1.1){$x_9$}
\put(5.1,1.1){$x_{11}$}
\put(2.1,2.1){$x_2$}
\put(3.1,2.1){$x_4$}
\put(4.1,2.1){$x_6$}
\put(5.1,2.1){$x_8$}
\put(3.1,3.1){$x_1$}
\put(4.1,3.1){$x_3$}
\put(5.1,3.1){$x_5$}
\end{picture}
\caption{\label{32} $(3,2)$-periodic initial value problem updated in direction
$(2,1)$.}
\end{center}
\end{figure}
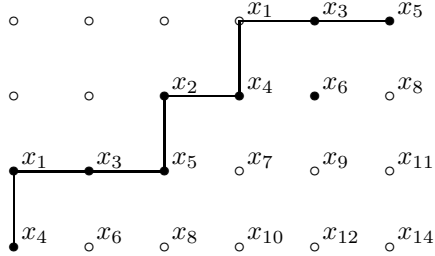

Next, we consider the mapping obtained from $\s$-periodic reduction taking
$s_1$ and $s_2$ to be coprime. Without loss of generality we may assume $s_1\geq s_2$.
Remember we denote $s_1+s_2=\q$. Initial values $x_1,x_2,\ldots,x_\q$ are given on
a standard staircase \cite{QCPN}, see also \cite{K} in which a general theory of periodic
reductions for equations not necessarily defined on a square has been developed.
The initial values are updated by an recurrence of order $\q$:
\begin{equation} \label{recxg}
x_{n+1}=P(x_{n-s_2},x_{n-s_1},x_{n-\q}),
\end{equation}
cf. equation (\ref{P}). For example, when $\s=(3,2)$ we pose initial values as in Figure
\ref{32}. These are updated by shifting over $(2,1)$, e.g. $x_5\mapsto x_6=P(x_4,x_3,x_1)$.

By setting $x_n=a_n/b_n$ we derive, for $n>\q$,
\begin{eqnarray}
a_n&=&a_{n-\q}w_n-\alpha b_{n-s_1}b_{n-s_2}b_{n-\q} \label{aa}\\
b_n&=&b_{n-\q}w_n \label{bb}
\end{eqnarray}
where $w_n=a_{n-s_1}b_{n-s_2}-a_{n-s_2}b_{n-s_1}$. We choose $b_i=1$, $i=1,2,\ldots,\q$
so that $a_n$ and $b_n$ are polynomials in $a_1,a_2,\ldots,a_\q$.
As before, from initial degrees $d^a_n=d^b_n+1=1$ ($n=1,2,\ldots,\q$) we find that
$d^a_n=d^b_n+1$ ($n\in\N$), and that
\[
d^a_n=d^a_{n-s_1}+d^a_{n-s_2}+d^a_{n-\q}-1, \quad d^b_n=d^b_{n-s_1}+d^b_{n-s_2}+d^b_{n-\q}+1
\]
are upper bounds for the degrees of $a_n$ and $b_n$. If $f^{t_k}$ divides $g_k$ with $k<n$,
then $f^{t_n}$ divides $g_n$, where
\begin{equation} \label{tn2}
t_n=t_{n-s_1}+t_{n-s_2}+t_{n-\q}.
\end{equation}
If initially $t_i=0$, $i=1,2,\ldots,\q-1$, $t_\q=2$ then
\[
f^2|g_n \Rightarrow f^{t_{\q+i}} | g_{n+i}, \ i\in\N.
\]

\begin{conjecture} \label{c1}
The polynomial $w_n^2$ divides $g_{n+\q}$ (for $n>\q$).
\end{conjecture}

It turns out this conjecture is more difficult to verify for $s_2<<s_1$.
We verified the conjecture in the following ranges of values $s_2<s_1$:
$s_2=1,\ldots,5$ with $s_2<s_1\leq 9s_2$, and $s_1=s_2+1$ with
$s_2=6,7,\ldots,25,50,100,150,200,250,1000$.

The conjecture would imply that $w_n^{t_i}$ divides $g_{n+i}$. Assuming it, we can define
polynomials $c_i$ by
\[
b_n=c_n \prod_{i=1}^{n-1} c_i^{t_{n-i}},
\]
which yields $d^b_n= d^c_n + (d^c \ast t)_n$.
Using the recurrences for $t$ and $d^b$ we find
\[
d^c_n=d^c_{n-s_1} + d^c_{n-s_2} - d^c_{n-\q} + 1.
\]
In the case $\s=(3,2)$ the sequence \cite[seq. A001399]{Slo}
\[
0, 0, 0, 0, 0, 1, 1, 2, 3, 4, 5, 7, 8, 10, 12, 14, 16, 19, 21, 24, \ldots,
\]
is given by
\[
d^c_n=\frac{47}{72} + \frac{(-1)^n}{8} + \frac{\zeta^n+\zeta^{-n}}{9} - \frac{1}{2}n+\frac{1}{12}n^2,\quad \zeta^3=1.
\]

In general, the quantity $d^a_n-d^g_n$ is bounded from above by $d^c_n+1$,
whose asymptotic growth is
\[
\sim (2s_1s_2)^{-1} n^2.
\]

\subsection{The exceptional case} \label{ExCa}
The case $\s=(3,1)$ is an exceptional case. Here the growth is linear,
which resembles the fact that that the mapping can be linearized.
Introducing $h=(x_1-x_3)(x_2-x_4)$, the mapping
\[
(x_1,x_2,x_3,x_4)\mapsto (x_2,x_3,x_4,x_1+\frac{\alpha}{x_4-x_2})
\]
reduces to $h\mapsto \alpha - h$, which is an involution.\footnote{The function $h$ is
a $2$-integral of the mapping. In \cite{KQ} $k$-symmetries are used to perform explicit
dimensional reduction of mappings related to $(s_1,1)$-periodic reductions of lattice KdV.
The dimension $s_1+1$ is reduced to $s_1$ or $s_1-2$, when $s_1$ is even or odd, respectively.}
Nevertheless, it is interesting to see what cancelations cause the growth to become linear.

We set $x_n=a_n/b_n$ to find
\begin{align}
a_n&=a_{n-4}(a_{n-3}b_{n-1}-a_{n-1}b_{n-3})-\alpha b_{n-1}b_{n-3}b_{n-4}, \label{ra}\\
b_n&=b_{n-4}(a_{n-3}b_{n-1}-a_{n-1}b_{n-3}). \label{rb}
\end{align}
Taking initial values $(a_1,a_2,a_3,a_4) = (x-w,y+z,-w,z)$ and $b_1=b_2=b_3=b_4=1$
we have found that, see the appendix,
\begin{align}
a_n&=y^{t_{n-2}}(\alpha-xy)^{t_{n-3}}x^{t_{n-4}}c_n, \label{ea}\\
b_n&=y^{s_{n+1}}(\alpha-xy)^{s_n}x^{s_{n-1}}, \label{eb}
\end{align}
where $t_0=t_1=s_0=s_1=0$ and
\begin{align}
t_{n+2}&=t_{n+1}+t_{n}+\lfloor \frac{n}{4} \rfloor, \label{tn22} \\
s_{n+2}&=s_{n+1}+s_{n}+(-1)^n \lfloor \frac{n}{4} \rfloor. \notag
\end{align}
Define $r_n=s_{n+4}-t_{n+1}$. One can show that $r_n=n(1+(-1)^n)/4$, which is nonnegative.
It follows that the $a_n/c_n$ is a common divisor of $a_n$ and $b_n$.
Dividing out this factor we are left with denominator growth ($n\geq 4$)
\[
d^b_n-d^{a/c}_n = r_{n-3}+2r_{n-4}+r_{n-5} = n - 4.
\]
Note that in this case the common divisor of $a_n$ and $b_n$ consists of
three different factors only, whereas for other values of $\s$ the number
of common divisors grows linearly with $n$. Here, the multiplicity grows
faster than what can be expected from the form of the recurrence. In other
word a `miracle' happens at every iterate: from (\ref{tn22}) one can derive
\[
t_{n+4}=t_{n+3}+t_{n+1}+t_{n}+\lfloor \frac{n}{2} \rfloor, \\
\]
which should be compared to (\ref{tn2}), taking $s_1=1,s_2=3,\q=4$.

\section{Growth of degrees of projective mappings}
The entropy of a rational mapping has also been defined in terms of
the growth of the degree of its equivalent in projective space \cite{BV}.
Again we first consider the case $\s=(2,1)$.
\subsection{A low dimensional example}
The 3-dimensional mapping is
\[
(x_1,x_2,x_3)\mapsto (x_2,x_3,x_1+\frac{\alpha}{x_3-x_2}).
\]
We set $x_i=a_i/a_4$, $i=1,2,3$. If we denote the image by $b_i/b_4$, then
the homogenised mapping is $a\mapsto b$:
\begin{equation} \label{he}
\left( \begin{array}{c}
a_1 \\ a_2 \\ a_3 \\ a_4
\end{array} \right) \mapsto \left( \begin{array}{c}
b_1 \\ b_2 \\ b_3 \\ b_4
\end{array} \right) = \left( \begin{array}{c}
a_2(a_3-a_2) \\ a_3(a_3-a_2) \\ a_1(a_3-a_2) + \alpha a_4^2 \\ a_4(a_3-a_2)
\end{array} \right) .
\end{equation}
Note that the first, second, and fourth component of the image share a common divisor.
We are interested in the growth of the multiplicities of such a divisor.
Suppose that $c$ divides $a_1,a_2$ and $a_4$. From (\ref{he}) it follows
that $c$ is a common divisor of $b_1,b_3,b_4$. We continue the argument,
\[
c|(a_1,a_3,a_4) \Rightarrow c|(b_2,b_3,b_4)
\]
and
\[
c|(a_2,a_3,a_4) \Rightarrow c^2|(b_1,b_2,b_4),\ c|b_3.
\]
However, if we denote the common divisor of $b_1,b_2,b_4$ by $c$, then
miraculously $c^3$ divides all four components of the fourth iterate of
$a\mapsto b$. At the next iterates the multiplicities double. Denoting
the multiplicity of $c$ in the fourth component of the $i$th iterate by $t_i$,
we have $t_1=t_2=t_3=1$, $t_4=3$, and (at least) $t_{n>4}=2t_{n-1}$.
We now introduce two sets of polynomials $c_i,d_i$ as follows:
\begin{eqnarray}
\left( \begin{array}{l}
a_1 \\ a_2 \\ a_3 \\ a_4
\end{array} \right) &\mapsto& \left( \begin{array}{l}
a_2c_1 \\ a_3c_1 \\ d_1 \\ a_4c_1
\end{array} \right) \mapsto \left( \begin{array}{l}
a_3c_1c_2 \\ d_1c_2 \\ d_2c_1 \\ a_4c_1c_2
\end{array} \right) \mapsto \left( \begin{array}{l}
d_1c_2c_3 \\ d_2c_1c_3 \\ d_3c_1c_2 \\ a_4c_1c_2c_3
\end{array} \right) \mapsto \notag \\
& & \left( \begin{array}{l}
d_2c_1^3c_3c_4 \\ d_3c_1^3c_2c_4 \\ d_4c_1^3c_2c_3 \\ a_4c_1^3c_2c_3c_4
\end{array} \right) \mapsto \left( \begin{array}{l}
d_3c_1^6c_2^3c_4c_5 \\ d_4c_1^6c_2^3c_3c_5 \\ d_5c_1^6c_2^3c_3c_4 \\ a_4c_1^6c_2^3c_3c_4c_5
\end{array} \right) \mapsto \cdots \mapsto \notag \\
& & \left( \begin{array}{l}
d_{n-2} \prod_{i=1}^{n-3} c_i^{t_{n+1-i}} c_{n-1}c_n \\
d_{n-1} \prod_{i=1}^{n-2} c_i^{t_{n+1-i}} c_n \\
d_{n} \prod_{i=1}^{n-1} c_i^{t_{n+1-i}} \\
a_4 \prod_{i=1}^{n} c_i^{t_{n+1-i}}
\end{array} \right) \mapsto \cdots . \label{nth}
\end{eqnarray}
As an ordinary polynomial map the degree of the $n$th iterate is
\[
2^n = 1 + (d^c \ast t)_{n+1}.
\]
Substracting $2^n=2+ 2(d^c \ast t)_{n}$ from this equation, and using the
recursion for $t$, we find
\[
d^c_n=d^c_{n-1}+d^c_{n-2}-d^c_{n-3}+1.
\]
Projectively, the $n$th iterate (with $n>2$) is
\[
(d_{n-2}c_{n-1}c_n, d_{n-1} c_{n-2} c_n, d_{n} c_{n-2}c_{n-1}, a_4 c_{n-2}c_{n-1}c_n),
\]
after division by the common factor $\prod_{i=1}^{n-3} c_i^{t_{n+1-i}}$. We define
\[
p_n:=a_4\prod_{i=\text{max}(1,n-3)}^{n-1} c_i.
\]
The projective degree is
\[
d^p_{n>3}=1+d^c_{n-1}+d^c_{n-2}+d^c_{n-3}.
\]
The recursion for $d^c$ yields $d^p_n=d^p_{n-1}+d^p_{n-2}-d^p_{n-3}+3$.
Together with initial values $d^p_i=2^{i-1}$, $i=1,2,3$ this gives
the sequence \cite[seq. A084684]{Slo}
\[
1,2,4,8,13,20,28,38,49,62, \ldots,
\]
which agrees with computations in projective space. The growth
\[
d^p_n=\frac{15}{8}+\frac{(-1)^n}{8}-\frac{3}{2}n+\frac{3}{4}n^2
\]
is the same as for a mapping connected to the discrete Painlev\'{e}
I equation \cite{BV,HV00}.

\subsection{More general periodic reductions}
Now we consider the projective mapping that corresponds to $\s$-periodic reduction
with $s_1$ and $s_2$ coprime. We take $s_1\leq s_2$, and $\q=s_1+s_2$.
It is convenient to take initial values $x_0,x_2,\ldots,x_{\q-1}$. They are updated
using the recurrence (\ref{recxg}), or equivalently, the $\q$-dimensional mapping
\[
(x_0,x_1,\ldots,x_{\q-1})\mapsto (x_1,\ldots,x_{\q-1},P(x_{s_1},x_{s_2},x_{0})).
\]
Denoting the image of $x_i=a_i/a_\q$ by $b_i/b_\q$ we find a mapping $a\mapsto b$
in $\q$-dimensional projective space,
\[
\left( \begin{array}{c}
a_0 \\ a_1 \\ \cdots \\ a_{\q-1} \\ a_\q
\end{array} \right) \mapsto \left( \begin{array}{c}
a_1(x_{s_1}-x_{s_2}) \\ a_2(x_{s_1}-x_{s_2}) \\ \cdots \\ a_{0}(x_{s_1}-x_{s_2})+a_\q^2  \\ a_\q (x_{s_1}-x_{s_2})
\end{array} \right) .
\]
As in the case $\s=(2,1)$ there is a common factor dividing all components but one.
When $s_1>1$ we have
\[
c|(a_0,a_1,\ldots, a_{\q-2},a_q) \Rightarrow \left\{ \begin{split}
c^2 &| (b_0,b_1,\ldots, b_{\q-3},b_{\q-1},b_q), \\ c &| b_{\q-2} \end{split} \right. .
\]
When $s_1>2$ we have
\[
\left\{ \begin{split} c^2 &| (a_0,a_1,\ldots, a_{\q-3},a_{\q-1},a_q), \\
c &| a_{\q-2} \end{split} \right. \Rightarrow
\left\{ \begin{split}
c^4 &| (b_0,b_1,\ldots, b_{\q-4},b_{\q-2},b_{\q-1},b_q), \\
c^3 &| b_{\q-3} \end{split} \right. .
\]
This doubling in most components continues until after $s_1-1$ iterations we are lead to (if $s_2>s_1+1$)
\[
\left\{ \begin{split} c^{2^{s_1-1}} &| (a_0,a_1,\ldots, a_{s_2-1},a_{s_2+1},\ldots,a_q), \\
c^{2^{s_1-1}-1} &| a_{s_2} \end{split} \right. \Rightarrow
\left\{ \begin{split} c^{2^{s_1}-1} &| (b_0,b_1,\ldots, b_{s_2-2},b_{s_2},\ldots,b_q), \\
c^{2^{s_1}-2} &| b_{s_2-1} \end{split} \right. .
\]
Then we have doubling again, until after $s_2-1$ iterations where the growth is similar to the above.
Doubling continues until ...
\begin{conjecture} \label{c2}
The `miracle' happens after $\q$ iterations where suddenly the multiplicity is one higher than double
the previous one.
\end{conjecture}
This we have only verified for a couple of small values of $s_1,s_2$. Conjecture \ref{c2} is harder
to verify, using direct calculation, than conjecture \ref{c1}. We will assume it in the sequel.
We define integer sequences by $t_1=1$ and
\[
t_{n+1}=\left\{ \begin{array}{ll}
2t_n-1,\quad &n=s_1,s_2, \\
2t_n+1,\quad &n=s_1+s_2, \\
2t_n,\quad &\text{otherwise.}
\end{array} \right.
\]
We now introduce two sets of polynomials $c_i,d_i$ as follows:
\begin{align}
\left( \begin{array}{l}
a_0 \\ a_1 \\ \vdots \\ a_{\q-2} \\ a_{\q-1} \\ a_\q
\end{array} \right) & \mapsto \left( \begin{array}{l}
a_1 c_1^{t_1} \\ a_2 c_1^{t_1} \\ \vdots \\ a_{\q-1} c_1^{t_1} \\ d_1 c_1^{t_1-1} \\ a_\q c_1^{t_1}
\end{array} \right) \mapsto \left( \begin{array}{l}
a_2 c_1^{t_2} c_2^{t_1} \\ a_3 c_1^{t_2} c_2^{t_1} \\ \vdots \\
d_1 c_1^{t_2-1} c_2^{t_1} \\ d_2 c_1^{t_2}c_2^{t_1-1} \\ a_\q c_1^{t_2} c_2^{t_1}
\end{array} \right) \mapsto \cdots \mapsto \notag \\
& \left( \begin{array}{l}
d_1 c_1^{t_\q-1} c_2^{t_{\q-1}} \cdots c_\q^{t_1} \\
d_2 c_1^{t_\q} c_2^{t_{\q-1}-1} \cdots c_\q^{t_1} \\
\vdots \\
d_{\q-1} c_1^{t_\q} \cdots c_{\q-1}^{t_2-1} c_\q^{t_1} \\
d_{\q} c_1^{t_\q} \cdots c_{\q-1}^{t_2} c_\q^{t_1-1} \\
a_\q c_1^{t_\q} \cdots c_{\q-1}^{t_2} c_\q^{t_1}
\end{array} \right) \mapsto  \left( \begin{array}{l}
d_2 c_1^{t_{\q+1}} c_2^{t_{\q}-1} \cdots c_\q^{t_1} \\
d_3 c_1^{t_{\q+1}} c_2^{t_{\q}} c_3^{t_{\q-1}-1} \cdots c_\q^{t_1} \\
\vdots \\
d_{\q} c_1^{t_{\q+1}} \cdots c_{\q}^{t_2-1} c_{\q+1}^{t_1} \\
d_{\q+1} c_1^{t_{\q+1}} \cdots c_{\q}^{t_2} c_{\q+1}^{t_1-1} \\
a_\q c_1^{t_{\q+1}} \cdots c_{\q}^{t_2} c_{\q+1}^{t_1}
\end{array} \right) \mapsto \cdots \mapsto \notag \\
& \left( \begin{array}{l}
d_{n-\q+1} \prod_{i=1}^n c_i^{t_{n+1-i}} / c_{n-\q+1} \\
d_{n-\q+2} \prod_{i=1}^n c_i^{t_{n+1-i}} / c_{n-\q+2} \\
\vdots \\
d_{n-1} \prod_{i=1}^n c_i^{t_{n+1-i}} / c_{n-1} \\
d_{n} \prod_{i=1}^n c_i^{t_{n+1-i}} / c_n\\
a_\q \prod_{i=1}^n c_i^{t_{n+1-i}}
\end{array} \right) \mapsto \cdots . \label{big}
\end{align}
As an ordinary polynomial map the degree of the $n$th iterate is
\[
2^n = 1 + (d^c \ast t)_{n+1}.
\]
Substracting $2^n=2+ 2(d^c \ast t)_{n}$ from this equation, and using the
recursion for $t$, we find
\[
d^c_n=\left\{ \begin{array}{ll}
1,\quad & 1\leq n\leq s_1, \\
d^c_{n-s_1} +1,\quad & s_1 < n \leq s_2, \\
d^c_{n-s_1} + d^c_{n-s_2} + 1,\quad & s_2 < n \leq \q, \\
d^c_{n-s_1} + d^c_{n-s_2} - d^c_{n-\q}+ 1,\quad & \q < n,
\end{array} \right.
\]
or, $d^c_n = d^c_{n-s_1} + d^c_{n-s_2} - d^c_{n-\q}+ 1$ for all $n$, taking $d^c_{n<1}=0$.

Projectively, the last component of the $(n-1)$st iterate is
\[
p_{n} := a_\q \prod_{i=\text{max}(1,n-\q)}^{n-1} c_i,
\]
which has degree
\[
d^p_n = 1+\sum_{i=\text{max}(1,n-\q)}^{n-1} d^c_i.
\]
We find
\begin{equation} \label{R}
d^p_n=\left\{ \begin{array}{ll}
n,\quad & 1\leq n\leq s_1+1, \\
d^p_{n-s_1} + n -1,\quad & s_1 +1 < n \leq s_2 +1, \\
d^p_{n-s_1} + d^p_{n-s_2} + n - 2,\quad & s_2 +1 < n \leq \q, \\
d^p_{n-s_1} + d^p_{n-s_2} - d^p_{n-\q} + \q,\quad & \q < n.
\end{array} \right.
\end{equation}
For example, in the case $\s=(2,3)$ the sequence of degrees
\[
1,2,3,5,8,12,16,22,28,35,43,52,61,72,83,95,108,122,136,\ldots
\]
is given by
\[
d^p_n=\frac{127}{72}+\frac{(-1)^n}{8}-\frac{\zeta^{n-1}+\zeta^{1-n}}{9}-\frac{5}{6}n+\frac{5}{12}n^2,\quad \zeta^3=1.
\]
In general, the recursion (\ref{R}) yields asymptotic growth
\[
\sim (s_1+s_2)(2s_1s_2)^{-1} n^2.
\]

\section{Conclusion}
In \cite{Viat} Viallet discussed two approaches: the {\it heuristic method},
where no proofs are obtained, and {\it serious singularity analysis}, which is
limited to 2-dimensional maps, or some exceptional higher dimensional cases.
The question was raised how can we go further, in particular to high dimensions?
One suggestion was given: the {\it arithmetical approach}, cf. \cite{Hal}.
In this paper we have presented a different approach and showed that it works
for high dimensions, at least for (most) mappings obtained as reductions from an
integrable lattice equation. The only condition on the dimension is that one has
to be able to iterate the $\q$-dimensional map $q$ times to verify conjecture
\ref{c1} or \ref{c2}. The scope of this approach is left open for future research,
e.g. to consider other reductions, other lattice equations, and non-integrable
or almost-integrable maps.

\appendix

\section{Solution of the (3,1)-map}
We prove that the recurrences (\ref{ra},\ref{rb}) yields expressions
(\ref{ea},\ref{eb}), with
\begin{equation} \label{ec}
c_{2n+1}=y(x-w)(xy)^{n-2} - P_{n}, \quad
c_{2n+2}=(z-y)(\alpha-xy)^{n-1}-yP_{n},
\end{equation}
where
\[
P_n:=\sum_{k=0}^{n-1} T^n_{n-k} (xy)^k \alpha^{n-1-k},
\]
with
\[
T^{n+1}_{k+1}=T^n_k-T^{n+1}_k,\quad T^n_0=T^n_n=1,
\]
that is, \cite[seq. A112468]{Slo}
\[
T^{n,k}=\sum_{i=k}^n (-1)^{n-i}{n+k-i-1 \choose n-i}.
\]

\noindent
{\bf Proof:} Substituting (\ref{ea},\ref{eb}) in (\ref{rb})
yields
\begin{equation} \label{rc}
c_{2n}=(\alpha-xy)c_{2n-2}-y(xy)^{n-2},\quad
c_{2n+1}=xyc_{2n-1}-(\alpha-xy)^{n-1}.
\end{equation}
Substituting (\ref{ec}) in (\ref{rc}) yields
\[
P_n=(\alpha-xy)P_{n-1}+(xy)^{n-1}, \quad
P_n=(xy)P_{n-1}+(\alpha-xy)^{n-1}
\]
which can be verified using the definition of $P$ and $T$.
Substituting (\ref{ea},\ref{eb}) in (\ref{ra})
yields
\[
(xy)^{i-4} y ( c_{2i+1} - (\alpha-xy)^{i-2} ) = - c_{2i-3}(c_{2i}-(\alpha-xy)c_{2i-2})
\]
and
\[
(\alpha-xy)^{i-4} ( c_{2i} - \alpha (xy)^{i-3} y) = - c_{2i-4}(c_{2i-1}-xy c_{2i-3}),
\]
which follow as a consequence of (\ref{rc}). \hfill $\square$

\vspace{2mm}
\noindent
{\bf Remark 1:}
The expressions for $x_n=a_n/b_n$ can be simplified as follows. Let
\begin{equation} \label{rex}
(x_1,x_2,x_3,x_4)=(x-w,y+z,-w,z),\quad x_{n>4}=x_{n-4}+\frac{\alpha}{x_{n-1}-x_{n-3}}.
\end{equation}
Then
\begin{equation} \label{eex}
x_{2n+1}=x - w - x \frac{P_n}{(xy)^{n-1}}, \quad
x_{2n+2}=y + z - y \frac{P_n}{(\alpha-xy)^{n-1}}.
\end{equation}

\vspace{2mm}
\noindent
{\bf Remark 2:}
The recursion (\ref{rex}) can be solved explicitly as follows. The variable
$y_n=x_n-x_{n+2}$ satisfies $(y_n+y_{n-2})y_{n-1}=\alpha$ and we find
\[
x_n=\left\{\begin{array}{ll}
x_1-\sum_{i=1}^{(n-1)/2} y_{2i-1}, \quad &n \text{ odd} \\
x_2-\sum_{i=1}^{n/2-1} y_{2i}, \quad &n \text{ even}
\end{array} \right.
\]
We can solve $y_n$ in terms of $z_n=y_ny_{n+1}$,
\[
y_n=\left\{\begin{array}{ll}
\frac{z_{n-1}}{z_{n-2}}\frac{z_{n-3}}{z_{n-4}}\cdots\frac{z_{2}}{z_{1}}y_1, \quad &n \text{ odd} \\
\frac{z_{n-1}}{z_{n-2}}\frac{z_{n-3}}{z_{n-4}}\cdots\frac{z_{3}}{z_{2}}y_2, \quad &n \text{ even}
\end{array} \right.
\]
and we have $z_n=\alpha-z_{n-1}$ which implies
\[
z_n=\left\{\begin{array}{ll}
z_1, \quad &n \text{ odd} \\
\alpha - z_1, \quad &n \text{ even}
\end{array} \right. ,
\]
where $z_1=y_1y_2=(x_1-x_3)(x_2-x_4)=xy$. Backsubstituting yields
\[
y_n=\left\{\begin{array}{ll}
\left(\frac{\alpha-xy}{xy}\right)^{(n-1)/2} x, \quad &n \text{ odd} \\
\left(\frac{xy}{\alpha-xy}\right)^{n/2-1} y, \quad &n \text{ even}
\end{array} \right.
\]
and we find (\ref{eex}) with
\begin{equation}\label{cpp}
P_n=\sum_{i=0}^{n-1}(\alpha-xy)^i(xy)^{n-i-1}=\frac{(\alpha-xy)^{n}-(xy)^{n}}{\alpha-2xy}.
\end{equation}

\vspace{2mm}
\noindent
{\bf Remark 3:}
From the latter expression (\ref{cpp}) for the polynomials $P_n$ it is easy to extract the following
properties
\begin{itemize}
\item[$\ast$] They form a divisibility sequence: $n|m \Rightarrow P_n|P_m$.
\item[$\ast$] The real parts of their zeros equal $\alpha/2$: $P_n(xy)=0 \Rightarrow xy + \bar{x}\bar{y} = \alpha$.
\end{itemize}

\subsection*{Acknowledgments} This research has been funded by the Australian Research Council
through the Centre of Excellence for Mathematics and Statistics of Complex Systems. I thank
Jarmo Hietarinta for introducing the heuristic approach at the SMS Summer School on Symmetries
and Integrability of Difference Equations, CRM (2008). The work was further developed during the
program Discrete Integrable Systems at the Isaac Newton Institute (2009) and I am grateful to its
hospitality. Thanks to Claude Viallet, Reinout Quispel and Dinh Tran for useful suggestions.


\begin{thebibliography}{10}
\bibitem{A}
V.I. Arnold, Dynamics of complexity of intersections,
Bol. Soc. Bras. Mat. 21 (1990), 1--10.

\bibitem{B}
M.P. Bellon, Algebraic Entropy of Birational Maps with Invariant Curves,
Lett. Math. Phys. 50 (1999), 79--90.

\bibitem{BMR}
S. Boukraa, J-M. Maillard, and G. Rollet, Integrable mappings and polynomial
growth, Physica A 209 (1994), 162--222.

\bibitem{BV}
M.P. Bellon and C.-M. Viallet, Algebraic entropy,
Comm. Math. Phys. 204 (1999), 425--437.

\bibitem{FV}
G.Falqui and C.-M. Viallet, Singularity, complexity and quasi-integrability of  rational mappings,
Comm. Math. Phys. 154 (1993), 111--125.

\bibitem{Hal}
R.G. Halburd, Diophantine integrability, J. Phys. A: Math. Gen. 38 (2005), 263–269.

\bibitem{HV98}
J. Hietarinta and C.-M. Viallet, Singularity confinement and chaos in discrete systems,
Phys. Rev. Let. 81 (1998), 326--328.

\bibitem{HV07}
J. Hietarinta and C.-M. Viallet, Searching for integrable lattice maps using factorization,
J.Phys. A 40 (2007), 12629--12643.

\bibitem{HV00}
J. Hietarinta and C.-M. Viallet, Discrete Painlev\'{e} I and singularity confinement in projective space,
Solitons and Fractals 11 (2000), 29--32.

\bibitem{K}
Peter H. van der Kamp, Initial value problems for lattice equations, J. Phys. A: Math. Theor. 42 (2009) 404019.

\bibitem{KQ}
Peter H. van der Kamp, G.R.W. Quispel, The staircase method, in preparation.

\bibitem{OTGR}
Y. Ohta, K. M. Tamizhmani, B. Grammaticos, and A. Ramani, Singularity confinement and algebraic entropy:
the case of the discrete Painlev\'{e} equations
Phys. Lett. A 262 (1999), 152--157.

\bibitem{QCPN}
G.R.W. Quispel, H.W. Capel, V.G. Papageorgiou, and F.W. Nijhoff, Integrable mapping derived from soliton equations,
Physica A 173 (1991), 243--266.

\bibitem{Slo}
\newblock N.J.A. Sloane, The On-Line Encyclopedia of Integer Sequences,
\newblock published electronically at www.research.att.com/$\sim$njas/sequences/.

\bibitem{TGR}
S. Tremblay, B. Grammaticos and A. Ramani, Integrable lattice equations and their growth properties,
Phys. Lett. A 278 (2001), 319--324.

\bibitem{Via}
C.-M. Viallet, Algebraic entropy for lattice equations, arXiv:math-ph/0609043.

\bibitem{Viat}
C.-M. Viallet, Algebraic entropy: measuring the complexity of rational dynamics, talk
given at Exeter (2005).

\bibitem{Ves}
A.P. Veselov, Growth and integrability in the dynamics of mappings,
Comm. Math. Phys. 145 (1992), 181-193.

\end{thebibliography}
\end{document}